\documentclass[aps,prl, amsmath, showpacs, preprintnumbers,superscriptaddress, twocolumn,sort&compress,floatfix, amssymb]{revtex4-1}
\usepackage{graphicx,color}
\usepackage{mathptmx, textcomp}
\usepackage[latin1]{inputenc}
\usepackage{braket}

\begin{document}

\title{Inducing Transport in a Dissipation-Free Lattice with Super Bloch Oscillations}

\author{Elmar Haller}
\author{Russell Hart}
\author{Manfred J. Mark}
\author{Johann G. Danzl}
\author{Lukas Reichs\"ollner}
\author{Hanns-Christoph N\"agerl}
\affiliation{Institut f\"ur Experimentalphysik and Zentrum f\"ur
Quantenphysik, Universit\"at Innsbruck, Technikerstra{\ss}e 25,
6020 Innsbruck, Austria}

\date{\today}

\begin{abstract}
Particles in a perfect lattice potential perform Bloch oscillations when subject to a constant force, leading to localization and preventing conductivity. For a weakly-interacting Bose-Einstein condensate (BEC) of Cs atoms, we observe giant center-of-mass oscillations in position space with a displacement across hundreds of lattice sites when we add a periodic modulation to the force near the Bloch frequency. We study the dependence of these ``super'' Bloch oscillations on lattice depth, modulation amplitude, and modulation frequency and show that they provide a means to induce linear transport in a dissipation-free lattice. Surprisingly, we find that, for an interacting quantum system, super Bloch oscillations strongly suppress the appearance of dynamical instabilities and, for our parameters, increase the phase-coherence time by more than a factor of hundred.
\end{abstract}

\maketitle

Understanding the conduction of electrons through solids is of fundamental concern within the physical sciences. The simplified situation of an electron under a constant force $F$ within a perfect, non-dissipative, periodic lattice was originally studied by Bloch and Zener \cite{Bloch1928} over 70 years ago. Their and subsequent studies revealed that the particle would undergo so-called Bloch oscillations (BOs), a periodic oscillation in position and momentum space, thereby quenching transport and hence resulting in zero conductivity. BOs can be viewed as periodic motion through the first Brillouin zone, resulting in a Bloch period $T_B=2 \hbar k/F$, where $k=\pi/d$ is the lattice wave vector for a lattice spacing $d$. They result from the interference of the particle's matter wave in the presence of the periodic lattice structure,
requiring a coherent evolution of the wave during the time $T_B$. Generally, it is believed that conductance is restored via dissipative effects such as scattering from lattice defects or lattice phonons \cite{Kanemitsu1995,Ashcroft1976}. In bulk crystals, relaxation processes destroy the coherence of the system even before a single Bloch cycle is completed. These systems thus exhibit conductivity but prevent the observation of BOs. To observe BOs, the BO frequency $\nu_B=1/T_B$ must be large compared to the rate of decoherence. In semiconductor superlattices, where the Bloch frequency is enhanced, a few cycles have been observed \cite{Leo1992}.

A recent approach to observe and study BOs is to use systems of ultracold atoms in optical lattice potentials with a force that is provided by gravity or by acceleration of the lattice potential. In these engineered potentials, generated by interfering laser waves, dissipation is essentially absent, and decoherence can be well-controlled \cite{Gustavsson2008b}. Essentially all relevant system parameters are tunable, e.g. lattice depth and spacing, particle interaction strength, and external force, i.e. lattice tilt. For sufficiently low temperatures, a well-defined, sufficiently narrow momentum distribution can initially be prepared. BOs have been observed for ultracold thermal samples \cite{Dahan1996,Battesti2004,Ferrari2006}, for atoms in weakly-interacting Bose-Einstein condensates (BECs)  \cite{Anderson1998,Morsch2001,Gustavsson2008b}, and for ensembles of non-interacting quantum-degenerate fermions \cite{Roati2004}. Non-interacting BECs \cite{Gustavsson2008,Fattori2008} are ideally suited to study BOs as interaction-induced dephasing effects are absent, allowing for the observation of more than $20000$ Bloch cycles \cite{Gustavsson2008}.

As for any oscillator, classical or quantum, it is natural that one investigates the properties of the oscillator under forced harmonic driving. In the context of ultracold atomic gases, this is readily possible, as $\nu_B$ is in the range of Hz to kHz. The dynamics of a
harmonically driven Bloch oscillator has recently been the subject of several theoretical \cite{Korsch2003,Hartmann2004,Thommen2002,Thommen2004,Kolovsky2009} and experimental studies \cite{Wilkinson1996,Sias2008,Ivanov2008,Alberti2009}. For example, modulation-enhanced tunneling between lattice sites \cite{Sias2008,Ivanov2008} and spatial breathing of incoherent atomic samples \cite{Alberti2009} have been observed. Here, for a weakly-interacting atomic BEC in a tilted lattice potential, we demonstrate that harmonic driving can lead to directed center-of-mass motion and hence to transport. In this dissipationless system we thereby recover transport and conductivity. More strikingly, for slightly  off-resonant driving, we observe giant matter-wave oscillations that extend over hundreds of lattice sites. These ``super Bloch oscillations'' result from a beat between the usual BOs and the drive. They are rescaled BOs in position space and can also be used, by appropriate switching of the detuning or the phase, to engineer transport. Interestingly, forced driving leads to strongly reduced interaction-induced dephasing and greatly extends the time over which ordinary BOs can be observed.

\begin{figure}[t]
 \includegraphics[width=8.5cm] {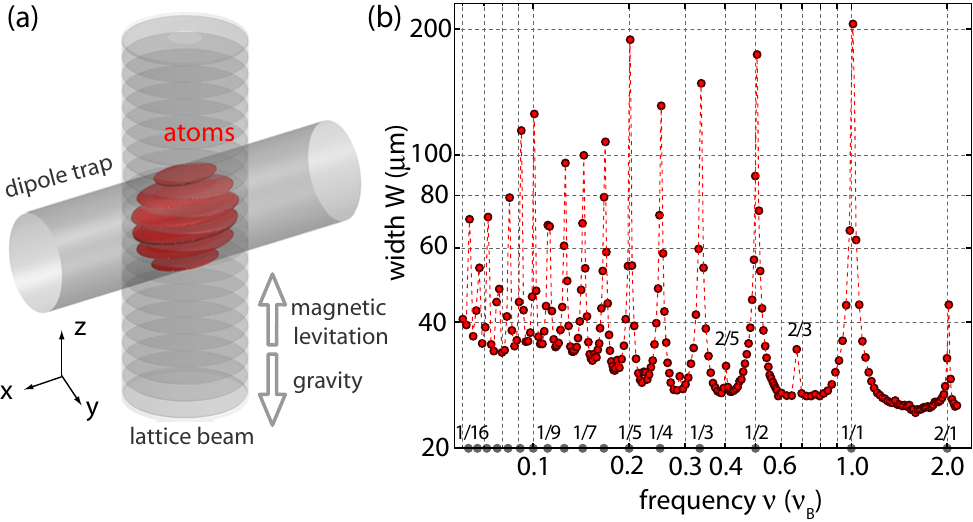}
  \caption{(color online) Experimental setup (a) and excitation spectrum (b) for atoms in a tilted periodic potential. The width $W$ is plotted as a function of the drive frequency $\nu$. The resonances correspond to a drastic spreading of the atomic wave packet as a result of modulation-assisted tunneling \cite{Sias2008} when $\nu \approx i/j \times \nu_B$, where $i,i$ are integers. The parameters are $F_0 = 0.096(1)mg$, $\Delta F=0.090(4)mg$, $V=3.0(3) \ E_R$, and $ \tau=2$ s. The dashed line is a guide to the eye.
 \label{fig1}}
\end{figure}

\begin{figure}[h]
 \includegraphics[width=8.5cm] {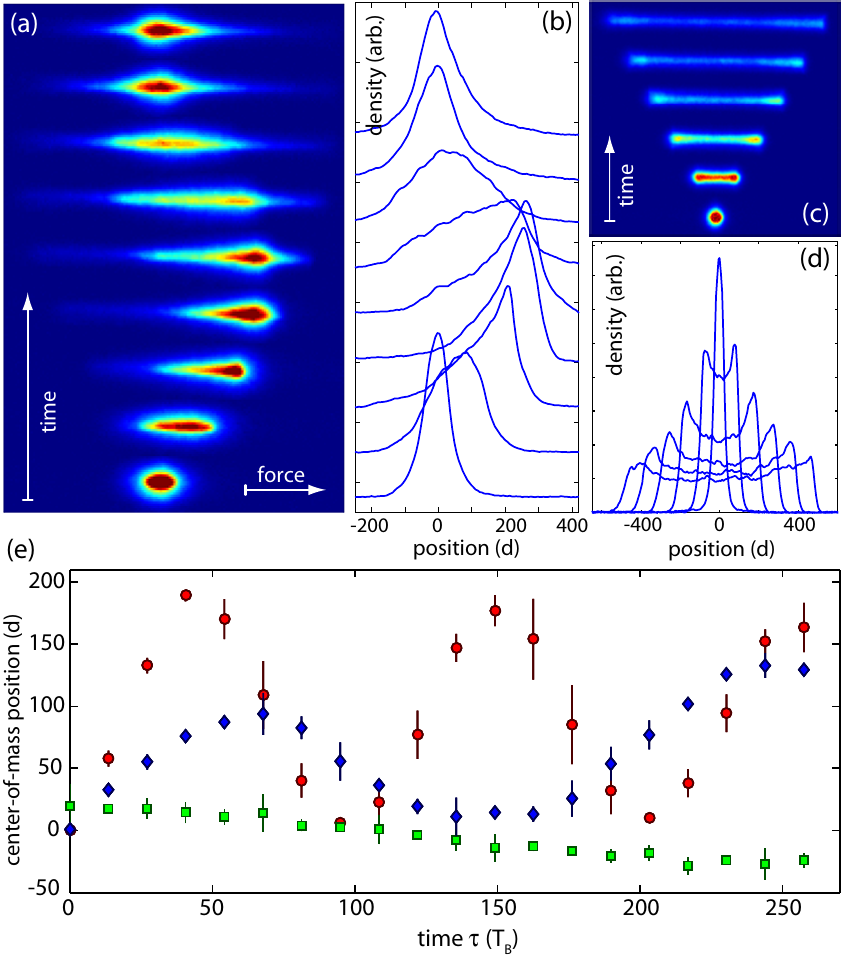}
 \caption{(color online) Observation of super Bloch oscillations and modulation-driven wave packet spreading. (a) and (b) In-situ absorption images and density profiles for off-resonant modulation ($\Delta \nu=-1$ Hz), showing giant oscillatory motion across more that 200 sites (time steps of 120 ms, average of 4 images). (c) and (d) In-situ absorption images and density profiles for resonant modulation ($\Delta \nu=0$ Hz), showing a wave packet that spreads symmetrically (time steps of $100$ ms, average of 4 images). For (a)-(d), the parameters are $F_0 = 0.062(1) mg$, $\Delta F = 0.092(4) mg$, $V=3.0(3)$$E_R$, $a=11(1)$ a$_0$. (e) Center-of-mass motion for $a=11(1) a_0$ (squares), $a=90(1) a_0$ (diamonds), $a=336(4) a_0$ (squares).
 \label{fig2}}
\end{figure}

The experimental starting point is a tunable BEC of $1.2\times10^5$ Cs atoms in a crossed beam dipole trap \cite{Kraemer2004} adiabatically loaded within  $400$ ms into a vertically oriented 1D optical lattice \cite{Gustavsson2008} as illustrated in Fig.~\ref{fig1}(a). The lattice spacing is $d=\lambda/2$, where $\lambda=1064.49(1)$ nm is the wavelength of the light. Unless stated otherwise, we work with a shallow lattice with depth $V=3.0(3)\ E_R$, where $E_R=h^{2}/(2 m \lambda^{2})$ is the photon recoil energy for particles with mass $m$. The atoms are initially levitated against gravity by means of a magnetic field gradient and spread across approximately $50$ lattice sites with an average density near $5\times10^{13} \ $cm$^{-3}$ in the central region of the sample. We control the strength of the interaction as measured by the s-wave scattering length $a$ near a Feshbach resonance \cite{Kraemer2004}. Throughout this work, unless stated otherwise, we work at $a=11(1)$ $a_0$, where $a_0$ is Bohr's radius. We initiate BOs by removing the dipole trap confinement in the vertical direction and by reducing the levitation in $1$ ms to cause a force that is a small fraction of the gravitational force $mg$, for which $\nu_B$ is near $100$ Hz. An additional harmonic modulation of the levitation gradient then results in an oscillating driving force $F(t) = F_0 + \Delta F \sin(2\pi\nu t + \phi)$, where $F_0$ is the constant force offset, $\Delta F$ is the amplitude of the modulation, $\nu$ is the modulation frequency, and $\phi$ is a phase difference between the BOs and the drive. After a given hold time $\tau$ we switch off all optical beams and magnetic fields and take in-situ absorption images after a short delay time of $800 \mu$s.

We first determine the excitation spectrum. Fig.~\ref{fig1}(b) shows the $1/\sqrt{e}$-width $W$ of the matter wave after $\tau=2$ s as a function of $\nu$. A series of narrow resonances at rational multiples of $\nu_B$ can clearly be identified. In agreement with recent experiments \cite{Sias2008,Ivanov2008}, we attribute these resonances to modulation-enhanced tunneling between lattice sites, leading to dramatic spreading of the atomic wave packet. Tunneling between nearest neighbor lattice sites is enhanced when $\nu_B$ is an integer multiple $j$ of $\nu$ via a $j$-phonon process \cite{Eckardt2005}, while tunneling between lattice sites $i$ lattice units apart is enhanced when $\nu$ is an integer multiple $i$ of $\nu_B$. Even combinations thereof, e.g. $i/j=2/3$ or $2/5$, are detectable.

We now investigate the dynamics of the wave packet in more detail. For this, we use the resonance with $i\!=\!j\!=\!1$ and choose $\nu=\nu_B+\Delta \nu$, where $\Delta \nu $ is the detuning. In Fig.~\ref{fig2}(a)-(d) we present absorption images and spatial profiles for the weakly-interacting BEC. The time evolution for the width, shape, and center position of the BEC is dramatic. On resonance ($ \Delta \nu = 0$), (c) and (d), the atomic ensemble spreads as it develops pronounced edges. We will see below that the center-of-mass motion depends crucially on the phase $\phi$. Off resonance, (a) and (b), for small detuning $ \Delta \nu = -1$ Hz, the wave packet exhibits giant oscillatory motion across hundreds of lattice sites that we denote as ``super Bloch oscillations'' (sBO). Note that, for the parameters used here, the amplitude for ordinary BOs corresponds to about $4d = 2.1 \ \mu$m. Also the width and higher moments of the distribution show oscillatory behavior. In Fig.~\ref{fig2}(e) we plot the center-of-mass position as a function of time for $\Delta \nu = -1$ Hz. At $a=11(1)$ $a_0$ we typically observe sBOs over the course of several seconds. The dynamics of sBOs strongly depends upon the site-to-site phase evolution of the matter-wave. In fact, stronger interactions, e.g. $a=90(1)$ $a_0$, distort the density profile of the driven BEC and alter the BEC's oscillation frequency and amplitude. For sufficiently strong interactions, no sBOs are observed. We also attribute the wave-packet spreading as seen after one cycle in Fig.~\ref{fig2}(b) mostly to interactions. For the measurements above, we intentionally use a large modulation amplitude $\Delta F$ to enhance the amplitude of sBOs. However, all effects equally exist for $\Delta F \ll F_0$, as we will also demonstrate below in Fig.~\ref{fig4}(b).

\begin{figure}[t]
 \includegraphics[width=8.5cm] {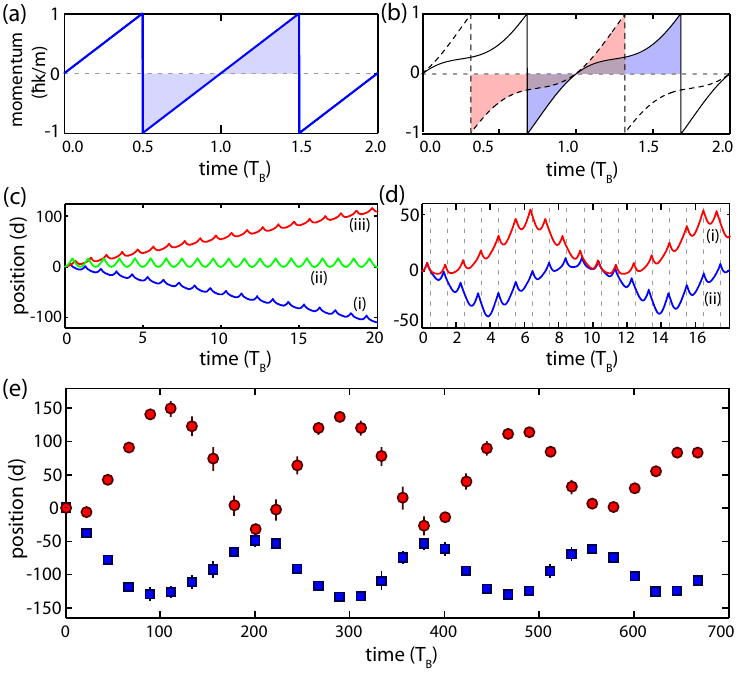}
  \caption{(color online) Results from a semi-classical model for sBOs. (a) For a constant force, here $F_0 = 0.06 mg$, the velocity (in units of $\hbar k/m$) exhibits a symmetric, saw-tooth-like time evolution, typical for BOs. (b) Resonant modulation, here with $\Delta F = 0.8 F_0$, alters the symmetric periodic velocity excursions of normal BOs ($\phi=0$, solid line, $\phi=\pi$, dashed line), leading to a net-movement, (c), with $\phi=0$ (i), $\phi=\pi/2$ (ii), and $\phi=\pi$ (iii). An additional detuning $\Delta\nu = \pm 0.1 \nu_B$ results in a periodically changing phase difference and hence in giant oscillations in position space, (i) and (ii) in (d). On top of the motion, normal BOs can clearly be seen. The phase of sBOs depends on the sign of $\Delta \nu$, as shown by experimental data in (e), where $F_0 = 0.096(1) mg$, $\Delta F = 0.090(4) mg$, $\Delta \nu = 1$ Hz (circles), $-1$ Hz (squares).
   \label{fig3}}
\end{figure}

\begin{figure}[t]
 \includegraphics[width=8.5cm] {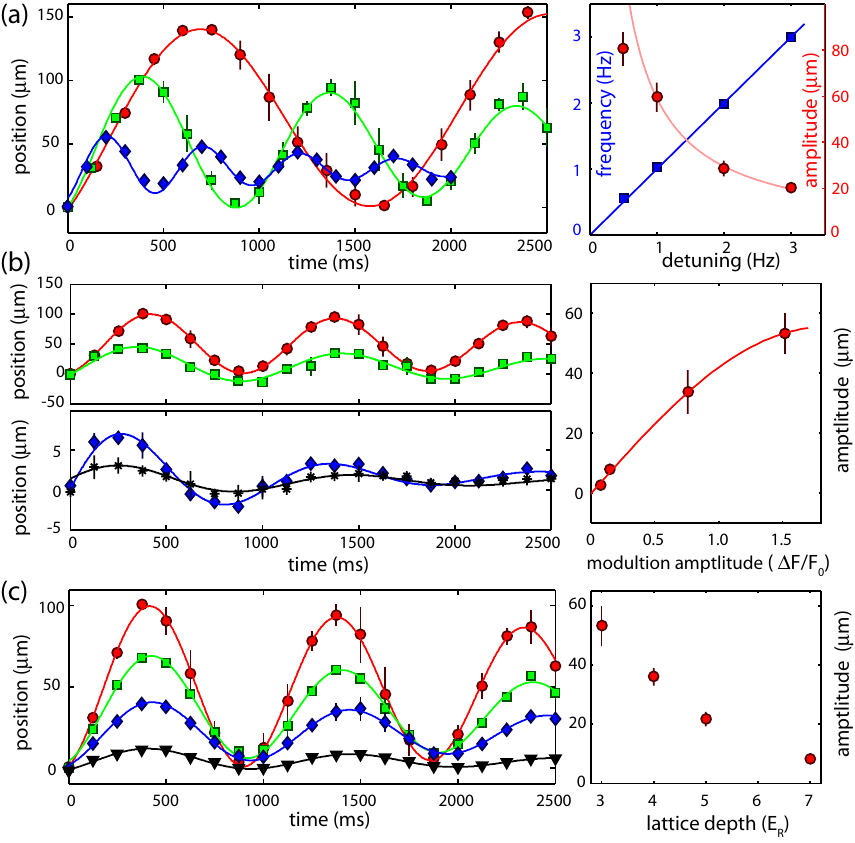}
 \caption{(color online) Quantitative analysis of sBOs. (a) The effect of the detuning $\Delta \nu$ on the oscillation frequency and the amplitude of sBOs, with $\Delta \nu=0.5$ Hz (circles), $1$ Hz (squares), $2$ Hz (diamonds). Right: The solid lines are fits with linear and $\Delta \nu^{-1}$-dependence, respectively. (b) Dependence of the amplitude of sBOs on $\Delta F/F_0$. The data sets correspond to $\Delta F/F_0=$ $1.52$ (circles), $0.76$ (squares), $0.15$ (diamonds), $0.08$ (stars). Right: The solid line is a fit proportional to $B_1(\Delta F/F_0)$. (c) Amplitude of sBOs as a function of lattice depth, V$=$ $3$ $E_R$ (circles), $4$ $E_R$ (squares), $5$ $E_R$ (diamonds), $7$ $E_R$ (stars). If not stated otherwise, the parameters for all measurements shown here are $F_0 = 0.062(1) mg$, $\Delta F = 0.092(4) mg$, $\Delta \nu = -1$ Hz.
  \label{fig4}}
\end{figure}

It is useful to develop a simple semi-classical model to obtain a qualitative understanding of the origin of sBOs. The only elements of this model are that the wave packet is accelerated by the applied force and that, once the wave packet reaches the edge of the first Brillouin zone, it is Bragg reflected. This model does not include an effective mass and cannot be used to predict quantitative results. Fig.~\ref{fig3}(a)-(d) shows the result of a numerical integration of the time-dependent acceleration $a(t) = F_0/m + \Delta F/m \sin(2\pi (\nu_B+\Delta \nu) t + \phi)$ with periodic Bragg reflection. For a constant acceleration $\Delta F=0$, the wave packet's velocity shows the well-known saw-tooth-like time evolution that corresponds to BOs. The curve in (a) is symmetric, hence, there is no net movement, as indicated by the shaded regions of equal area. If, however, there is additional harmonic modulation at $\nu=\nu_B$, the velocity excursions will not be symmetric about zero, (b), and result in a net movement for each period, leading to linear motion, (c). Only for $\phi=\pi/2$ or $\phi=3\pi/2$ symmetry is restored and no net movement will occur. Note that, in general, the velocity of the linear motion depends non-trivially on $\phi$. Off-resonant modulation with $\Delta \nu \ll \nu_B $ induces a slowly-varying phase mismatch between the drive and the original Bloch period. This results in a slow oscillation of the net movement for each Bloch cycle, which finally sums up to a giant oscillation in position space, (d). Evidently, this oscillation is the result of a beat between the drive and the original BO. The initial direction of the motion depends on $\phi$ and $\Delta \nu$. In particular, a change in the sign of $\Delta \nu$ at a given $ \phi $ can lead to opposite motion in position space, as verified experimentally in Fig.~\ref{fig3}(e) for $\Delta \nu =\pm 1$ Hz.

A quantitative understanding of sBOs \cite{Kolovsky2009} can be obtained from an approach based on Wannier-Stark states \cite{Thommen2002,Thommen2004}. In essence, the harmonic drive is expected to lead to a rescaling of the tunneling rate $J \to J_\mathrm{eff}=J B_1(\Delta F/F_0)$ and the force $F_0 \to F_\mathrm{eff} = h \Delta \nu/d$ for a stationary lattice with tilt. Here, $B_1$ is the first Bessel function of the first kind. The amplitude of sBOs is thus given by a new Wannier-Stark localization length $L_\mathrm{eff} = J_\mathrm{eff}/(d F_\mathrm{eff} )$ \cite{Kolovsky2009}. In this sense, sBOs are rescaled BOs. We quantitatively study the dependence of amplitude and period of sBOs on $\Delta\nu$, $\Delta F/F_0$, and $V$. The results are shown in Fig.~\ref{fig4}. As expected, the period $T$ is given by $1/\Delta \nu$. Also, the oscillation amplitude scales as $1/\Delta \nu$, and its Bessel-function dependence on $\Delta F/F_0$ is well reproduced. Given our spatial resolution, we can observe sBOs down to $\Delta F/F_0 = 0.08$ (Fig.~\ref{fig4}(b)). Note that sBOs can only be observed with sufficient wave function coherence and for well-defined initial conditions, i.e. for sufficient wave packet localization in the first Brillouin zone of the lattice. Nevertheless, incoherent atomic samples exhibit a breathing of the spatial distribution \cite{Alberti2009} as the oscillation period is insensitive to the initial conditions. In the work of Ref.\cite{Alberti2009}, the breathing can be understood in terms of an incoherent sum over localized Wannier-Stark states that individually show a breathing motion with period $T$ \cite{Thommen2004}.

\begin{figure}[t]
 \includegraphics[width=8.5cm] {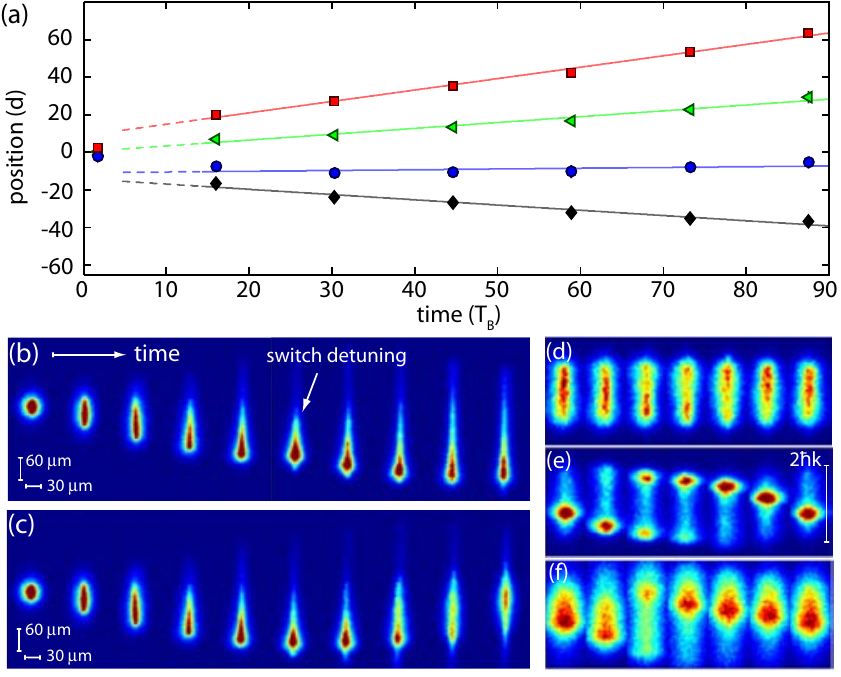}
 \caption{(color online) Inducing transport and suppressing inter\-action-induced dephasing. (a) Linear motion for resonant modulation. $\Delta \phi=0^{\circ}$ diamonds, $65^{\circ}$ circles, $120^{\circ}$ triangles, $190^{\circ}$ squares. $\Delta \phi=0^{\circ}$ and $\Delta \phi=190^{\circ}$ were chosen to maximize the speed in opposite directions. The solid lines are linear fits to the data points excluding the first data point. (b) Directed motion for off-resonant modulation. $\Delta\nu$ was switched from $ -1 $ Hz to $ 1 $ Hz after $400$ ms. For comparison, (c) shows the oscillatory motion without switching (time steps of 80 ms). The parameters are $F_0 = 0.096(1) mg$, $\Delta F = 0.090(4) mg$. (d)-(f) BOs in quasi-momentum space after 10 BO cycles with no drive (d), after 17 BO cycles with drive (e), after 750 BO cycles with drive (f). The parameters are $F_0 = 0.096(1) mg$, $\Delta F =  0.045(3) mg$, $\Delta\nu = -1 \ $ Hz. The time steps are 1 ms.
   \label{fig5}}
\end{figure}

The results above provide two mechanisms to circumvent the localization inherent in BOs and to induce coherent transport in an otherwise insulating context. As shown in Fig.~\ref{fig5}(a), resonant modulation ($\Delta\nu=0$) causes directed motion of the wave packet's center-of-mass. For longer times, we find that the motion is approximately linear. The mean velocity depends on the relative phase $\phi$ of the Bloch oscillator and the drive. In the experiment, we varied $\phi$ via $\phi=\phi_0 + \Delta \phi$, where $\phi_0$ is a constant phase offset, which depends on the details how BOs are initiated. For off-resonant modulation, transport can be induced by switching the sign of $\Delta\nu$ before a half-cycle of a sBO is completed. The wave packet then continues to move in the original direction. This motion is shown in Fig.~\ref{fig5}(b), where we switch the sign after $400$ ms. For comparison, Fig.~\ref{fig5}(c) shows a sBO with $T=1$ s without switching.

Surprisingly, we find that harmonic modulation strongly reduces the effect of dephasing and wave-packet spreading in quasi-momentum space as a result of interactions. Fig.~\ref{fig5}(d)-(f) shows single Bloch cycles in quasi-momentum space. Without modulation, the wave packet spreads across the first Brillouin zone within about $10$ Bloch cycles. This phenomenon is well known in the context of interacting BECs and is attributed to the appearance of dynamical instabilities \cite{Wu2003}. In contrast, with modulation, the quasi-momentum distribution remains narrow and BOs can still be well resolved after $750$ Bloch cycles. Evidently, the effects of dynamical instabilities are strongly suppressed.

In summary, we have studied the coherent evolution of matter waves in tilted periodic potentials under forced driving and have observed giant sBOs, which result from a beat of BOs with the drive when a small detuning $\Delta\nu$ from the Bloch frequency is introduced. Localization as a result of BOs is broken, allowing us to engineer matter wave transport over macroscopic distances in lattice potentials with high relevance to atom interferometry \cite{Cronin2009}. We are now in a position to investigate the effect of interactions on driven transport, for which subdiffusive and chaotic dynamics have been proposed \cite{Kolovsky2009b}.

During the final preparation of the manuscript we became aware of related work on non-dissipative transport in a quantum ratchet \cite{Salger2009}. We thank A. R. Kolovsky, A. Zenesini, and A. Wacker for discussions and R. Grimm for generous support. We gratefully acknowledge funding by the Austrian Ministry of Science and Research and the Austrian Science Fund and by the European Union within the framework of the EuroQUASAR collective research project QuDeGPM. R.H. is supported by a Marie Curie International Incoming Fellowship within FP7.

\bibliographystyle{apsrev}

\end{document}